\documentclass[longauth]{aa}  

\usepackage{graphicx}
\usepackage{txfonts}
\usepackage{hyperref}
\usepackage{natbib}
\usepackage{float}
\usepackage[usenames,dvipsnames]{color}
\usepackage{enumitem}
\usepackage{amsmath}
\usepackage{amssymb}
\usepackage{changes}
\usepackage{academicons}
\usepackage{xcolor}
\usepackage{subfigure}
\usepackage{placeins}
\usepackage{booktabs}
\usepackage{ulem}
\usepackage[switch]{lineno}

\definecolor{darkblue}{cmyk}{55,17,0,0}
\hypersetup{colorlinks=true,
  linkcolor=darkblue,
  linkbordercolor=darkblue,
  citecolor=darkblue,
  urlcolor=black
  }

\bibpunct{(}{)}{;}{a}{}{,}

\newcommand{\sophi}{{\sc SO/PHI}}
\newcommand{\hmi}{{\sc SDO/HMI}}
\newcommand{\so}{{SO}}

\newcommand{\blos}{$B_{\rm LOS}$}
\newcommand{\blosOverMu}{$B_{\rm LOS}/\mu$}
\newcommand{\ic}{$I_{\rm c}$}
\newcommand{\au}{AU}

\begin{document} 

   \title{Intensity contrast of solar network and faculae close to the solar limb, observed from two vantage points}

   \author{K.~Albert\inst{1}\orcid{0000-0002-3776-9548}\thanks{\hbox{Corresponding author: K. Albert.} \hbox{\email{albert@mps.mpg.de}}}
     \and
   N.~A.~Krivova\inst{1}\orcid{0000-0002-1377-3067} \and
   J.~Hirzberger\inst{1} \and
   S.~K.~Solanki\inst{1}\orcid{0000-0002-3418-8449} \and
   A.~Moreno~Vacas\inst{2}\orcid{0000-0002-7336-0926} \and
   D.~Orozco~Su\'arez\inst{2}\orcid{0000-0001-8829-1938} \and
   N.~Albelo~Jorge\inst{1}\and
   T.~Appourchaux\inst{3}\orcid{0000-0002-1790-1951} \and 
   A.~Alvarez-Herrero\inst{4}\orcid{0000-0001-9228-3412} \and
   J.~Blanco Rodr\'iguez\inst{5}\orcid{0000-0001-9228-3412} \and
   A.~Gandorfer\inst{1}\orcid{0000-0002-9972-9840} \and
   P.~Gutierrez-Marques\inst{1}\orcid{0000-0003-2797-0392} \and
   F.~Kahil\inst{1}\orcid{0000-0002-4796-9527} \and
   M.~Kolleck\inst{1} \and
   R.~Volkmer\inst{6} \and
   J.C.~del~Toro~Iniesta\inst{2}\orcid{0000-0002-3387-026X} \and
   J.~Woch\inst{1}\orcid{0000-0001-5833-3738} \and 
   B.~Fiethe\inst{7}\orcid{0000-0002-7915-6723} \and
   I.~P\'erez-Grande\inst{9}\orcid{0000-0002-7145-2835} \and 
   E.~Sanchis~Kilders\inst{5}\orcid{0000-0002-4208-3575} \and
   M.~Balaguer~Jiménez\inst{2}\orcid{0000-0003-4738-7727} \and
   L.R.~Bellot~Rubio\inst{2}\orcid{0000-0001-8669-8857} \and
   D.~Calchetti\inst{1}\orcid{0000-0003-2755-5295} \and
   M.~Carmona\inst{8}\orcid{0000-0001-8019-2476} \and
   A.~Feller\inst{1} \and
   G.~Fernandez-Rico\inst{1,9}\orcid{0000-0002-4792-1144} \and
   A.~Fern\'andez-Medina\inst{4}\orcid{0000-0002-1232-4315} \and
   P.~Garc\'ia~Parejo\inst{4}\orcid{0000-0003-1556-9411} \and 
   J.L.~Gasent~Blesa\inst{5}\orcid{0000-0002-1225-4177} \and 
   L.~Gizon\inst{1,10} \orcid{0000-0001-7696-8665}
   B.~Grauf\inst{1} \and 
   K.~Heerlein\inst{1} \and
   A.~Korpi-Lagg\inst{1}\orcid{0000-0003-1459-7074} \and
   T.~Lange\inst{7} \and 
   A.~L\'opez Jim\'enez\inst{2} \and 
   T.~Maue\inst{6,11} \and 
   R.~Meller\inst{1} \and
   R.~M\"uller\inst{1} \and
   E.~Nakai\inst{6} \and 
   W.~Schmidt\inst{6} \and
   J.~Schou\inst{1}\orcid{0000-0002-2391-6156} \and
   J.~Sinjan \inst{1}\orcid{0000-0002-5387-636X} \and
   J.~Staub\inst{1}\orcid{0000-0001-9358-5834} \and
   H.~Strecker \inst{2}\orcid{0000-0003-1483-4535} \and 
   I.~Torralbo\inst{9}\orcid{0000-0001-9272-6439} \and 
   G.~Valori\inst{1}\orcid{0000-0001-7809-0067} 
   }
 
   \institute{
         Max-Planck-Institut f\"ur Sonnensystemforschung, Justus-von-Liebig-Weg 3,
         37077 G\"ottingen, Germany \\ \email{solanki@mps.mpg.de}
         \and
         Instituto de Astrofísica de Andalucía (IAA-CSIC), Apartado de Correos 3004,
         E-18080 Granada, Spain \\ \email{jti@iaa.es}
         \and
         Univ. Paris-Sud, Institut d’Astrophysique Spatiale, UMR 8617,
         CNRS, B\^ atiment 121, 91405 Orsay Cedex, France
         \and
         Instituto Nacional de T\' ecnica Aeroespacial, Carretera de
         Ajalvir, km 4, E-28850 Torrej\' on de Ardoz, Spain
         \and
         Universitat de Val\`encia, Catedr\'atico Jos\'e Beltr\'an 2, E-46980 Paterna-Valencia, Spain
         \and
         Leibniz-Institut für Sonnenphysik, Sch\" oneckstr. 6, D-79104 Freiburg, Germany
         \and
         Institut f\"ur Datentechnik und Kommunikationsnetze der TU
         Braunschweig, Hans-Sommer-Str. 66, 38106 Braunschweig,
         Germany
         \and
         University of Barcelona, Department of Electronics, Carrer de Mart\'\i\ i Franqu\`es, 1 - 11, 08028 Barcelona, Spain
         \and
         Instituto Universitario "Ignacio da Riva", Universidad Polit\'ecnica de Madrid, IDR/UPM, Plaza Cardenal Cisneros 3, E-28040 Madrid, Spain
         \and
         Institut f\"ur Astrophysik, Georg-August-Universit\"at G\"ottingen, Friedrich-Hund-Platz 1, 37077 G\"ottingen, Germany
         \and
         Fraunhofer Institute for High-Speed Dynamics,
         Ernst-Mach-Institut, EMI, Ernst-Zermelo-Str. 4, 79104 Freiburg, Germany
         }

\date{Received December 31, 2018; accepted January 1, 2019}

 
  \abstract
   {The brightness of faculae and network depends on 
   the angle at which they are observed
   and the magnetic flux density. 
   Close to the limb, assessment of this relationship has until now been hindered by the increasingly lower signal in magnetograms.}
   {This preliminary study aims at highlighting the potential of using simultaneous observations from different vantage points to better determine the properties of faculae close to the limb.}
   {We use data from the Solar Orbiter/Polarimetric and Helioseismic Imager (\sophi{}), and the Solar Dynamics Observatory/Helioseismic and Magnetic Imager (\hmi{}), recorded at $\sim60^\circ$ angular separation of their lines of sight at the Sun. We use continuum intensity observed close to the limb by \sophi{} and complement it with the co-observed \blos{} from \hmi{}, originating closer to disc centre (as seen by \hmi), thus avoiding the degradation of the magnetic field signal near the limb.}
   {We derived the dependence of facular brightness in the continuum on disc position and magnetic flux density from the combined observations of \sophi{} and \hmi{}. Compared with a single point of view, we were able to obtain contrast values reaching closer to the limb and to lower field strengths. We find the general dependence of the limb distance at which the contrast is maximum on the flux density to be at large in line with single viewpoint observations, in that the higher the flux density is, the closer the turning point lies to the limb. There is a tendency, however, for the maximum to be reached closer to the limb when determined from two vantage points.
   We note that due to the preliminary nature of this study, these results must be taken with caution.}
   {Our analysis shows that studies involving two viewpoints can significantly improve the detection of faculae near the solar limb and the determination of their brightness contrast relative to the quiet Sun.}

    
    \titlerunning{Intensity contrast of network and faculae from two vantage points}

\maketitle
%
%


\section{Introduction}
The solar photospheric magnetic field is organised into the mainly weak-field low-lying loops forming the internetwork (found predominantly in the quiet Sun), which is almost invisible in white light, and kG-strength magnetic flux tubes that manifest themselves as dark sunspots and pores and bright faculae and network \citep[see, e.g.][]{Solanki2006MagneticField}. The brightness of a given flux tube (also often referred to as a magnetic element) depends on its size and the angle to the observer.

Horizontal pressure balance with the environment leads to an evacuation of the magnetic flux-tube interior, in order to maintain hydrostatic equilibrium in the presence of magnetic pressure.
Therefore, for rays roughly parallel to the axis of the flux tube, the observable layer at optical depth $\tau = 1$ lies deeper than the surrounding photosphere.
Whereas inside the flux tube, the magnetic field inhibits convective energy transport, 
the walls of the flux tubes appear bright due to heating from the surrounding convection, the so-called “hot wall effect” \citep[see, e.g.][]{spruit76}. The balance between the lateral radiative heating and magnetic suppression of convection within a given flux tube depends on its diameter. Flux tubes tend to be close to solar surface normal due to magnetic buoyancy \citep[see][]{Buehler2015Plage, Jafarzadeh2014Incl}. Therefore, the position of a flux tube on the observed solar disc determines which part of the flux tube we see, as a consequence of the angle at which we observe it. 
Going from the disc centre to the limb, the hot walls of the flux tubes rotate into view, before the edge closer to the observer starts to obscure the opposite, observer-facing wall \citep[for extensive discussions, see][]{solanki_smallscale_1993, carlsson_observational_2004, keller_origin_2004}. 

The small-scale flux tubes forming faculae and network are typically not resolved by full-disc magnetographs, while the kG flux tubes in the internetwork have so far only been resolved in exceptional cases, for example by \citet{Lagg2010FullyResolved} who used the IMaX magnetograph on the Sunrise balloon-borne solar observatory  \citep[e.g.,][]{Solanki2010Sunrise, Barthol2011Sunrise, MartinezPillet2011IMAX}.
Thus, the size of faculae and network elements is difficult to assess. Instead, the magnetic flux density within the resolution element (pixel) is often used to describe the fraction of the solar surface covered by strong fields (often referred to as the magnetic filling factor). The intrinsic field strength of the flux tubes forming network and faculae is roughly unchanged, and their average size increases with the magnetic flux density. Therefore, it is also an indirect measure of the size of the magnetic features. 

Determining the relationship between facular and network brightness and the magnetic flux density as well as the distance of the faculae from the solar limb is important for our understanding and modelling of the radiant properties and thermal structure of faculae and the magnetic elements they are composed of. This also provides a classic constraint to magnetohydrodynamic simulations \citep[see][]{Beeck2015SmallScale} and is relevant for studying and modelling the variations of solar irradiance at time scales of days to millennia, which is driven by the intensity excess created by faculae and intensity deficit resulting from sunspots \citep[see][]{krivova-2003, solanki-2013, shapiro_nature_2017, yeo_publishers_2017, yeo_dimmest_2020}. In this study, we treat the facular and network features without differentiation, and refer to them as faculae collectively \citep{SolankiStenflo1984PropFluxtubes}. \citet{Criscuoli2017PhotometryNetwFac_HMI_PSF} and \citet{Buehler2019Comparison} showed benefits of treating them separately \citep[see also][]{Foukal2011}, and this would be interesting to address in a future study building on the current one.

Numerous studies examined the intensity contrast of faculae (i.e. their intensity relative to that of the internetwork) in relation to their magnetic flux density \citep[see][and references therein]{kobel_continuum_2011, kahil_2019}, as well as their distance from disc centre \citep[see][]{ortiz2002Intensity, Yeo2013Intensity}. For the magnetic flux density, most studies used the line-of-sight (LOS) magnetic field (\blos{}) observations, which is by far the most reliable of the magnetic components obtained from data utilising the Zeeman effect. 

However, analysis of the facular contrast becomes complicated and very uncertain close to the solar limb for multiple reasons.
Firstly, foreshortening effects become critical.
Thus, since most of the observed magnetic field is nearly vertical, the \blos{} component becomes weak close to the limb leading to a low signal-to-noise ratio, which is further reduced due to lower light levels (limb darkening). In addition, the spatial resolution of observations is reduced, when approaching the limb, with each pixel representing an increasingly larger surface area on the Sun due to foreshortening. Within a pixel with larger coverage, the contribution of a given spatially unresolved magnetic element to the magnetogram signal and to the contrast of that pixel is smaller. Furthermore, the chances of opposite-polarity flux cancellation within such pixels are higher, resulting in lower magnetic flux density measurements.

Secondly, the $\mu$-value of the observation influences the observed height: closer to the limb, the absorption line in which the measurements are taken, forms higher in the atmosphere than close to the disc centre \citep[see][for a discussion]{Schou23}. Whereas this change in observation height is part of what we aim to observe in the intensity, it is an undesired effect in the case of \blos{} measurements, which will be different to those carried out closer to the disc centre, distorting any comparisons between the two.

Thirdly, the incident polarised radiation from the Sun also depends on the angle of the observation, owing to the properties of the Zeeman effect. Due to radiative transfer effects and the finite width and geometry of solar magnetic features, the Stokes~$V$ amplitude need not scale linearly with $\mu$ \citep[see][]{Solanki1998ReliabilityStokes}. These additional effects include, but are not limited to: (1) changes in the width and strength (including potential saturation) of the spectral line as a consequence of the lower temperature and increased turbulent velocities sensed by the line towards the limb. (2) Possible changes in the Zeeman saturation of the Stokes V profile due to changes in field strength as a consequence of greater formation height of the line and the larger inclination of the field relative to the LOS. (3) The passage of individual rays through both magnetised and unmagnetised gas, etc.

Fourthly, the identification and isolation of facular features is also more challenging close to the limb because the apparent extension of the magnetic canopy of sunspots increases towards the limb  \citep[]{GiovanelliJones1982ThreeD, Solanki1994infrared7}. As the horizontal field of the canopy give a large contribution to Stokes $V$, it can be mistaken for a facular contribution \citep[see][]{Yeo2013Intensity, Ball2012TSIReconstr}.

Because of these restrictions, studies of the contrast of bright magnetic features are typically curtailed near the solar limb. For example, \citet{Yeo2013Intensity} studied facular intensity contrast as a function of distance from the disc centre (in terms of the cosine of the heliocentric angle, denoted as $\cos \theta = \mu$) and the measured magnetic flux density, normalised by $\mu$ (\blosOverMu{}), using data from the Helioseismic and Magnetic Imager on-board the Solar Dynamics Observatory \citep[SDO/HMI, see][]{schou_design_2012}. (The normalisation by $\mu$ corrects for the geometrical effects to first order under the assumption that the field is vertical to the local solar surface.) Due to the factors discussed above, weak network features ($B_{\text{LOS}}/\mu<50$\,G) could not be identified in the magnetograms near the limb ($\mu \leq 0.4$), so that the centre-to-limb variation (CLV) of facular intensity contrast in this regime remains unclear.

These restrictions could be overcome (or at least their severity reduced) if, in addition to a magnetograph on the ground, or in Earth orbit, a second such instrument observing the Sun from a different viewpoint were available. 
The Solar Orbiter mission \citep[SO or SolO;][]{muller_solar_2020} brings this new perspective to solar observations: it reaches a wide range of positions outside the Sun-Earth line and it carries the Polarimetric and Helioseismic Imager \citep[SO/PHI;][]{solanki_polarimetric_2020}, the first solar magnetograph to provide data with significant angular separations from Earth.
A combined analysis of simultaneous observations by \sophi{} and on-ground or Earth orbiting instruments, presents an opportunity to examine the Sun from two perspectives simultaneously. In particular, this allows substituting  the uncertain \blos{} measurements at the limb with more certain ones, inferred from observations closer to the disc centre, hence improving on earlier studies of facular contrast.

In this study, we present such an effort. To highlight the potential of such multi-angle studies in better constraining the dependence of facular brightness on the measured magnetic flux density and the distance to the limb, we combine simultaneous observations from \sophi{} and \hmi{} \citep{schou_design_2012}. We stress that we do not aim to provide final results. Instead, this study demonstrates that such a combination of viewpoints can indeed improve our knowledge of facular contrast, in particular close to the solar limb.

The paper is structured as follows. We present our method in Sections~\ref{Sec:Data&Proc} and \ref{Sec:DataCombination} by first detailing the observations from \sophi{} and \hmi{} and their processing, followed by describing how we combine the two vantage points. In Sect.~\ref{Sec:Results}, we derive and discuss the relationship of the facular contrast to $\mu$ and \blosOverMu{} from the combined data obtained by the two instruments, and from \sophi{}'s perspective alone. In Sect.~\ref{Sec:Conclusions} we summarise our findings and discuss how the obtained results can be improved and extended in the future.
   
\begin{figure}
   \centering
   \resizebox{\hsize}{!}
            {\includegraphics[]{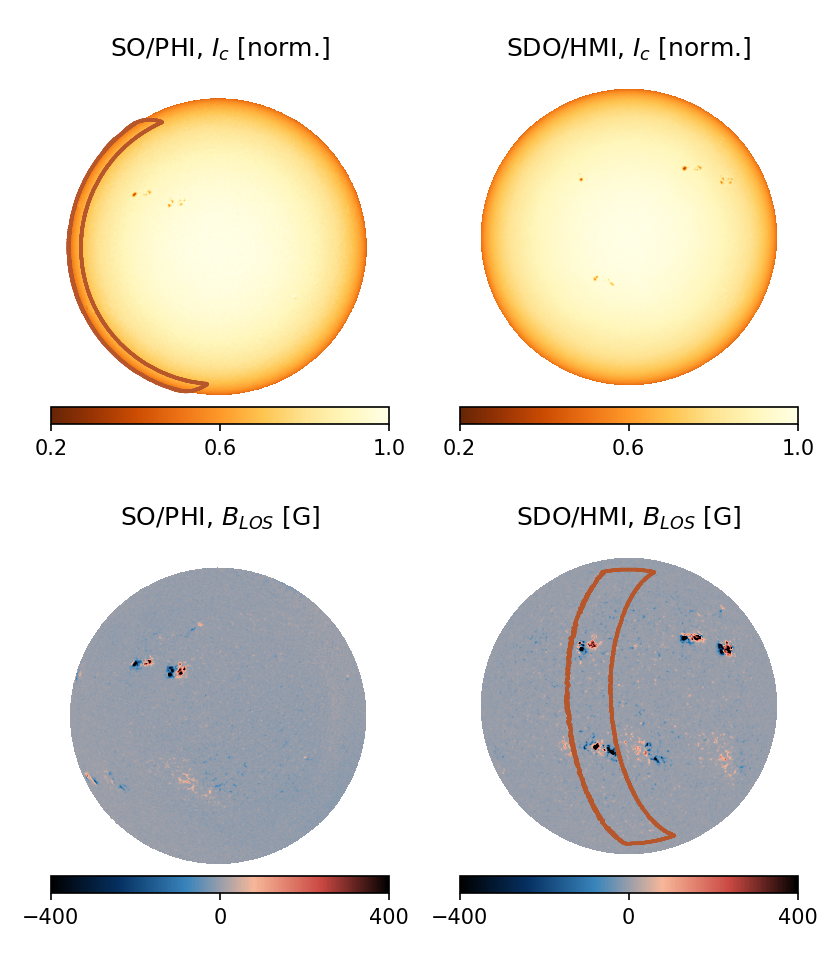}}
      \caption{The \ic{} (upper row) and \blos{} (lower row) observed by \sophi{} full disc telescope (FDT, left column) and \hmi{} (right column), on 6. Sept. 2021. The two instruments observed the Sun with $59.45^\circ$ angular separation. The brick-red outlines on the panels show the regions that we combined for the analysis: \ic{} from \sophi{} and \blos{} from \hmi{} within the overlap region of the two instruments, with $0.1<\mu_{\text{SO/PHI}}<0.4$ and $\mu_{\text{SDO/HMI}}>0.4$.
              }
         \label{Fig:Data}
   \end{figure}

\section{Data and their processing}\label{Sec:Data&Proc}

\subsection{Data}

The \sophi{} \citep{solanki_polarimetric_2020} is an imaging spectropolarimeter, sampling the photospheric \ion{Fe}{I}\,617.3\,nm absorption line. It has two telescopes: the Full Disc Telescope (FDT) and the High Resolution Telescope \citep[HRT, see][]{gandorfer18}. As their names suggest, the FDT covers the full solar disc at all phases of the spacecraft's orbit (with a pixel plate scale $3.75"$), while the HRT only images a fraction of it (pixel plate scale $0.50"$).
The \sophi{} measures the full Stokes vector ($I$, $Q$, $U$ and $V$) at six wavelength positions: five inside the spectral line, and one in the nearby continuum. From these, through the Zeeman and Doppler effects, the vector magnetic field ($\vec{B}$) and the LOS velocity ($v_{\rm LOS}$) can be determined at the average formation height of the spectral line. In addition, the continuum intensity is also returned. In this study, we use data products obtained in the longitudinal mode of the \sophi{} instrument. This is a simplified mode, which is applied on-board to reduce processing time and telemetry volume \citep[see][]{albert_autonomous_2020}. It provides the continuum intensity (\ic{}), and \blos{} (instead of $\vec{B}$), calculated with analytical formulae via the centre-of-gravity technique \citep{Semel1967Contribution, Rees1979LineFormation, Landi2004Polarisation}. These results are referred to as classical estimates. The measurement duration for a full data set is approximately $33$\,s.

The \hmi{} observes the same absorption line, using the same principle. Some relevant differences between the instruments are the pixel plate scale ($0.505"$ for \hmi{}), the sampling wavelengths of the line (\hmi{}'s six wavelength samples are uniformly spaced and centred over the line, sometimes resulting in the continuum not being directly sampled), and while it provides similar data products, it uses somewhat different techniques to derive them. Here we use the $720$-second data products: the reconstruction of continuum intensity, and the LOS magnetogram (\verb|hmi.M_720s|; calculated with the MDI-like algorithm, see \citealt[][]{couvidat_observables_2016}).

We use 10 pairs of \sophi{} -- \hmi{} observations, recorded during the cruise phase of \so{}, one from each day in the period 1 to 10 September 2021. The \sophi{} data used in the study is of the full solar disc, recorded with the FDT. During the observations, the angular separation of the two instruments changed from $67^\circ$ to $52^\circ$, while SO's distance to the Sun varied from $0.58$\,\au{} to $0.60$\,\au{}.
The \sophi{}-FDT at these distances observes the Sun with a radius of 440 to 453 pixels. The \sophi{} data have been fully reduced on-board the spacecraft, including the calibration and the determination of \blos{} \citep[for details of the on-board processing see][]{albert_autonomous_2020}. Due to the early phase of the mission, and the novelty of the on-board data processing, the calibration data and processes applied during the reduction of these data sets are preliminary.

Since we are mainly interested in extending earlier studies of facular contrast to locations closer to the solar limb, we analyse areas that appear at $0.1 < \mu < 0.4$ in \sophi{} data, and at $\mu > 0.4$ in \hmi{} data. As an example, Fig.~\ref{Fig:Data} shows \ic{} and \blos{} from co-observations of \sophi{} and \hmi{} on 6 September 2021. The regions that we analyse lie within the brick-red contours: the \ic{} at the limb from \sophi{}, and the corresponding area in the \hmi{} \blos{} at large $\mu$ values. We remark that combining the data the other way around, that is taking \ic{} from \hmi{} (from the limb), and complementing them with \blos{} from \sophi{} (closer to disc centre) would also be possible. However, such a combination is expected to be less accurate, as we have higher noise levels in the \sophi{} magnetograms (due to e.g. the lower amount of temporal averaging). 

\subsection{Attuning the observations}

To prepare the \hmi{} data products for combination with \sophi{} data, we convolve them with the point spread function (PSF) of the \sophi{}-FDT. As results of more accurate studies were not yet available, we used a theoretical PSF: the Airy disc of the telescope. This assumes a perfect telescope, only limited by the diffraction of light. To arrive at the effective PSF, we adjust this to the difference in distance to the Sun of the two instruments. Due to the large difference in aperture size (140\,mm in \hmi{} vs. 17.5\,mm in \sophi{}-FDT) which is not nearly compensated by the difference in distance to the Sun ($1$\,\au{} for \hmi{} and $0.6$\,\au{} for \sophi{}), we consider the \hmi{} PSF to be negligible relative to that of \sophi{}. 
More accurate PSF estimates, available now \citep[see][]{Bailen2023PSF, Kahil2023_PSF}, will be used in subsequent studies. 

Next, we resample the \hmi{} data to match the spatial sampling of the \sophi{}-FDT observations (i.e. we bin the \hmi{} data by a non-integer factor). We achieve this in the Fourier domain. We crop the convolved data (conserving the lower frequency regions) to a dimension which after the inverse transform will provide the same solar radius in pixels as we observe in the corresponding \sophi{} data set.

For our analysis, the \blos{} values come exclusively from \hmi{} observations (degraded and resampled to mimic \sophi{}), while the intensity contrast values are exclusively from \sophi{}. Thus, in principle, we do not need to worry about the magnetic and continuum intensity cross-calibration of the two instruments. However, the cross-calibration might have an effect on the comparison of the results obtained from combining the two viewpoints with those obtained from \sophi{} only, and can therefore affect our results presented in Fig.~\ref{Fig:curvesNewMask_mu} (see Sect.~\ref{Sec:Results}). We see a good continuity of the results, therefore believe that for the scope of this pre-study we can use data that has not been cross-calibrated.
Efforts to cross-calibrate \hmi{} and \sophi{}-FDT data products are underway (2022, priv. comm. with A. Moreno Vacas), and should be used by future studies.

\subsection{Identification of faculae}\label{Sec:Identif}

\begin{figure*}[tbp]
   \centering
    \begin{minipage}[c]{\hsize}
    \centering
    
    \includegraphics[width=.49\hsize]{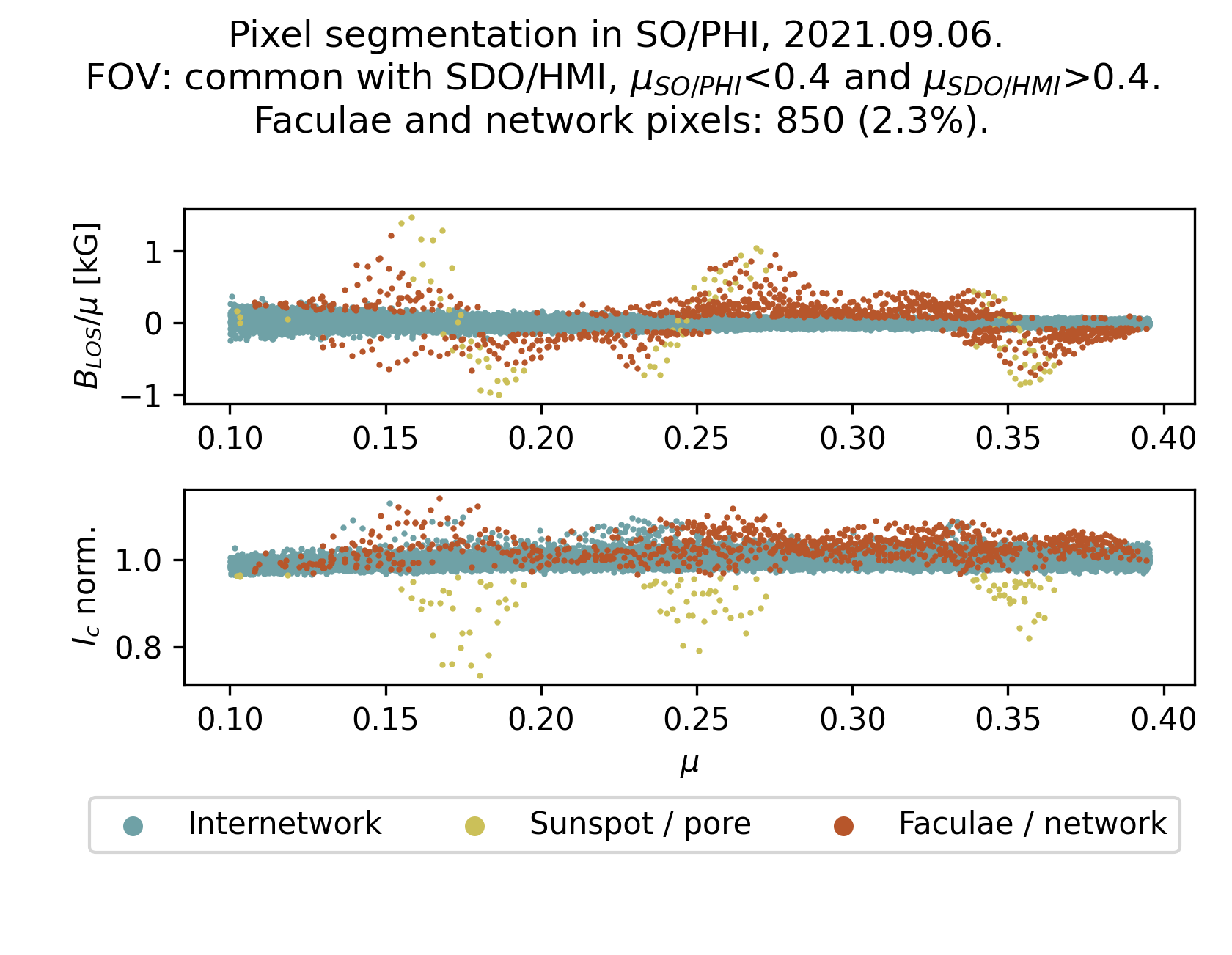}
    \includegraphics[width=.49\hsize]{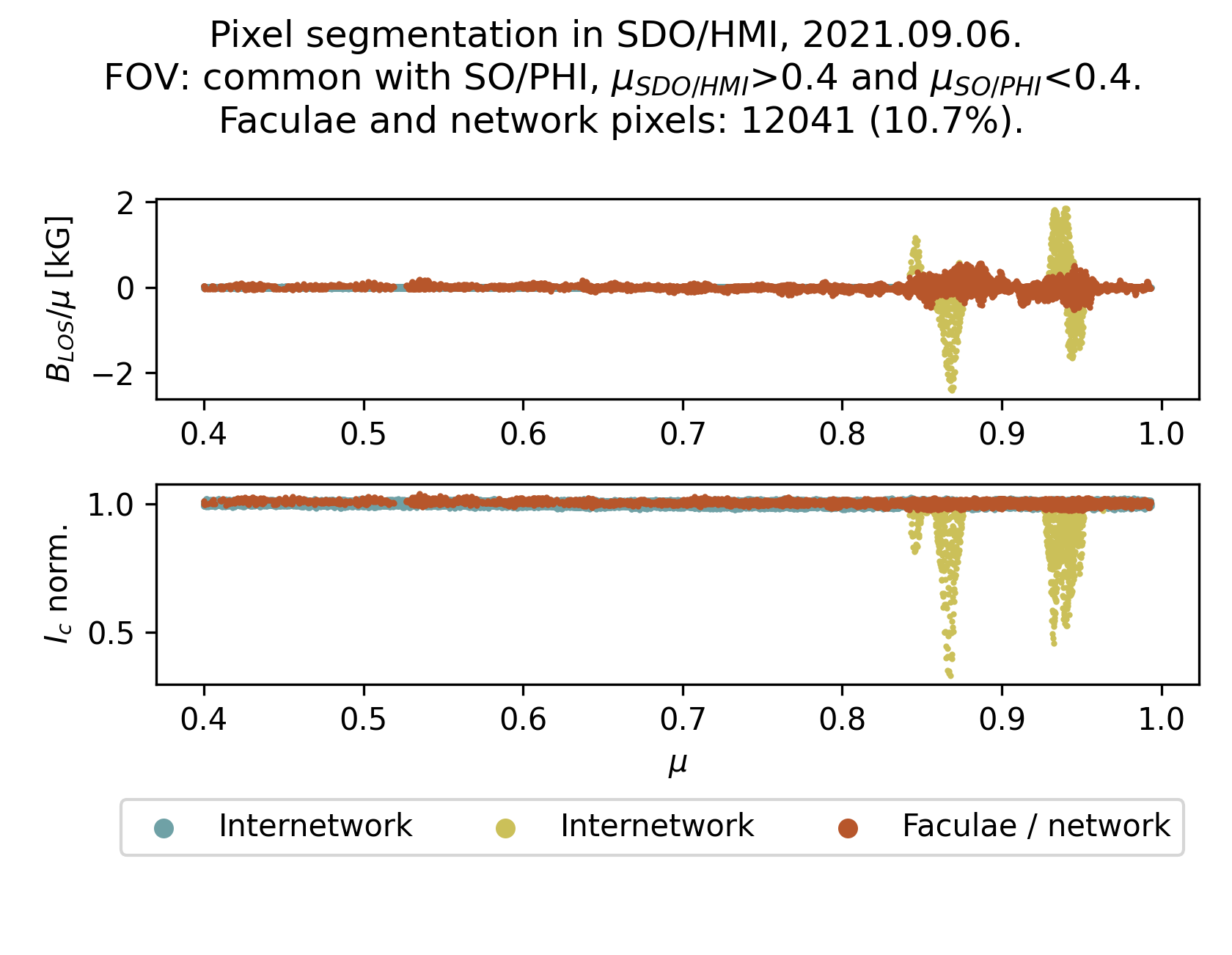}
    \end{minipage}

     \caption{Classification of pixels within the common area of interest in both instruments ($0.1 < \mu_{SO/PHI} \leq 0.4$ and $\mu_{SDO/HMI} > 0.4$) and their distribution with $\mu$, from images recorded on 6 Sept. 2021. In this field of view, we identify $2.2\%$ of the \sophi{} pixels (close to limb) and  $6.2\%$ of \hmi{} pixels (closer to disc centre) as faculae.
     }
         \label{Fig:SegmentationExample}
   \end{figure*}
   
We identify faculae/network, sunspots/pores, and internetwork following the method described by \citet{Yeo2013Intensity} \citep[for a discussion on the effects of identification methods, see][]{Centrone2003CLVFac_ident}. This is done for both instruments individually. In the case of \hmi{}, we use the PSF-degraded and resampled data. Faculae and network are identified in the magnetograms by their elevated \blos{} levels, while sunspots and pores are identified in the \ic{} images by lower intensity levels. Finally, parts of the Sun that do not qualify to be either network/faculae or sunspots/pores are counted as internetwork.

We first detect pixels with magnetic signals sufficiently above the noise level of the \blos{} maps:
\begin{equation}
B_{\text{LOS}}(x,y) > 3 \sigma_{B_{\text{LOS}}}(x,y), 
\label{Eq:Segm_Blos}
\end{equation}
where $x$ and $y$ are detector plane coordinates,
\blos{} is the measured line-of-sight magnetic field, and $\sigma_{B_{\text{LOS}}}$ is the standard deviation of \blos{}, a measure of the noise in \blos.

We determined $\sigma_{B_{\text{LOS}}}$ following the method described by \cite{ortiz2002Intensity} and \cite{Yeo2013Intensity}. This method includes two steps: (1) a computation of the CLV of the noise, and (2) a computation of the deviations from the obtained noise CLV profile. 
Thus, we first calculated the standard deviation, $\sigma$, of the magnetogram signal over concentric rings of pixels at similar distances from the disc centre and removed outliers, outside $3\sigma$, iteratively, until convergence. We also fitted a polynomial to the standard deviation versus distance profiles, thus establishing the CLV noise profiles.
Afterwards, we used a moving window to determine any deviations from the CLV noise profile, and fitted a polynomial surface to the result. We note, that if a window contains mostly active region pixels, the standard deviation within it is higher than for a quiet Sun region.
Therefore, our noise levels should rather be considered as an upper limit.
We prefer this conservative approach, where we may miss some facular pixels, over the potential inclusion of some noise in the analysis.

When we apply this method to the $720$\,s \hmi{} magnetograms at their original resolution, our approach returns the noise level $4$ to $7$\,G. This is close to the values of $4.7$ to $7.8$\,G reported by \cite{Liu2012CompBLOS}. We note that the \hmi{} team used more accurate methods to determine the noise level than we do.
After changes in the processing of the $720$\,s magnetograms in 2016, these values are expected to be somewhat lower (2022, priv. comm. with Yang Liu). Through analysing internetwork pixels, \citet{KorpiLagg2022QuietSunMagn} found $\sim 5.5$\,G noise levels prior to 2016, and $\sim 4.75$\,G afterwards.

For the resampled \hmi{} magnetograms, we find the noise level in the range from $3.5$ to $5.2$\,G. 
The downsampling of the data by nearly a factor of four, that is averaging over $\sim 16$ pixels, does not reduce the noise level by a factor of four. This reduction would be expected if the photon noise were the only noise source, however, that
is not the case for the 720\,s HMI LOS magnetograms \citep[see][]{Liu2012CompBLOS}.
Moreover, at the resulting resolution, while quiet Sun weak magnetic fields are visible, they are wrongly classified as noise.
This again leads to a conservative pixel classification, as we potentially exclude some pixels from the analysis which harbour network magnetic fields.

In the case of the \sophi{} magnetograms, the variation of the noise level over the field of view does not show a clear CLV. Instead, it is dominated by a large-scale gradient across the field of view, the origin of which is still under investigation. This indicates that also for \sophi{}, the noise level is not driven by photon noise only. 
In the absence of a clear CLV of the noise profile for the \sophi{} magnetograms, we calculate the noise directly with the moving window. This yields noise levels ranging from $7.1$\,G to $10.4$\,G, which is higher than what we find in the resampled \hmi{} data.

Next, we identified the pixels that belong to sunspots or pores based on \ic{}.
To achieve this, we first calculated the CLV of the quiet Sun at the continuum wavelength following the method described by \cite{neckel_solar_1994}. We then found the large-scale deviations of the quiet Sun \ic{} from the CLV, which we consider to be a residual of the flat field correction, following \citet{Yeo2013Intensity}. The \sophi{} \ic{} observations are subject to a ghost image, which is a faint ($\sim 0.5\%$ of the intensity) duplicate image of the Sun overlaid on the solar disc with a small spatial offset. Therefore, we first mitigated the effect of this ghost image on the sensor, and then determined the residuals on the result.

We normalised \ic{} by its CLV and the flat field residuals:
\begin{equation}\label{Eq:Icnorm}
I_{\rm c, norm}(x,y,t) = \frac{I_{\rm c}(x,y,t)}{{\rm CLV}_{I_{\rm c}}(x,y)  R_{I_{\rm c}}(x,y)},
\end{equation}
where $t$ is time (which refers to one of the 10 considered data sets), $I_{c,\text{norm}}$ denotes the normalised \ic{}, $\text{CLV}_{I_\text{c}}$ denotes the centre to limb variation of the quiet Sun intensity, and $R_{I_\text{c}}$ marks residuals of the flat field correction, present in the \ic{} data.

To obtain an intensity threshold for identifying sunspots, we again followed \citet{Yeo2013Intensity}. For the 10 data sets, we derived the quiet Sun $I_{\rm c, norm}$ standard deviation, which we denote $\rm \sigma_{I_{\rm c},\rm QS}$. The threshold separating sunspots from the internetwork ($I_{\rm c, threshold}$) was set, conservatively, at the mean of the minimum value of $I_{c,\text{norm}}(x,y,t) - 3  \rm \sigma_{I_{\rm c},\rm QS}(x,y,t)$ for each of the 10 days.
The $I_{\rm c, threshold}$ is $0.965$ in \sophi{}, and $0.975$ in \hmi{}. We consider all pixels below these values to belong to sunspots or pores. 

As a final step in our pixel segmentation, we found all isolated pixels identified as faculae based on the previous criteria. We considered these to be false positives, and therefore treated them as internetwork fields.

Figure~\ref{Fig:SegmentationExample} shows the distribution of the pixels of interest derived from the magnetograms of both instruments. The top panels show \blosOverMu{} at various $\mu$ values, while the bottom panels show \ic{}. The \sophi{} $I_c$ (bottom left panel) shows a weak downward trend when approaching the limb, indicating a bias in the normalisation of \ic{}. This is a result of imprecision in determining the radius of the Sun in the images, mainly due to two factors: low spatial resolution, and the as yet uncorrected distortion of the \sophi{}-FDT. In \sophi{} magnetograms, where the region was close to the limb, we find 853 facular pixels, which is $2.2\%$ of all pixels in this area. In the resampled \hmi{} data, where the region appeared closer to the disc centre, we identify 12041 pixels as faculae, that is $10.7\%$ of all pixels in this area. The difference in the fraction of pixels identified as faculae is due to the combined effect of three factors.
(1) The noise level in the \hmi{} data is lower than that in \sophi{}.
(2) The signal-to-noise ratio of the \blos{} measured close to the disc centre is higher than that measured closer to the limb. 
(3) Due to the foreshortening effect, pixels close to the limb represent a larger surface area on the Sun, which leads to more averaging and thus more potential cancellation of oppositely signed magnetic flux (in SO/PHI data) as compared to pixels recorded closer to the disc centre (HMI data).

The produced maps that mark the locations of the facular pixels were then used for our analysis (see Sect.~\ref{Sec:Results}). 
   
\subsection{Definition of the intensity contrast}\label{Sec:IntCont}

Following \cite{Yeo2013Intensity}, we calculate the continuum intensity contrast of the facular pixels, $C_{I_\text{c}}$, as:
\begin{equation}
    C_{I_\text{c}}(x,y,t) = I_{c,\text{norm}}(x,y,t) - 1,
    \label{Eq:Int_contr}
\end{equation}
where $I_{c,\text{norm}}(x,y,t)$ is defined in Sect.~\ref{Sec:Identif}.

\section{Combining the two vantage points} \label{Sec:DataCombination}

Our goal here is to assign to each \ic{} pixel measured by \sophi{} at the limb the corresponding \blosOverMu{} value measured by \hmi{} closer to the disc centre. The close-to-centre \hmi{} measurements are a “super-sampled” version of what \sophi{} measured close to limb: due to foreshortening, several pixels at a large $\mu$ value combine into a single one at the limb (i.e. the spatial coverage of the pixels increases). Thus, to assign the \sophi{} limb pixels with the \blosOverMu{} values measured by \hmi{}, we first need to cross-match the pixels in the observations by the two instruments, i.e., we have to find which \blosOverMu{} pixels in \hmi{} correspond to the \ic{} pixels measured by \sophi{}. To do this co-alignment, we re-projected
the \hmi{} \blosOverMu{} data to the coordinate system of \sophi{}, using the SunPy \citep[see][]{the_sunpy_community_sunpy_2020} implementation of the method described in \citet{deforest_re-sampling_2004}. This method uses a Hanning window to weigh the input pixels in the footprint of each output pixel, reducing aliasing effects, and producing values close to the mean.
Thereby, the mean of \blosOverMu{} over the input pixels is (roughly) preserved in the output pixel.
We additionally correct for the time difference in the data acquisition of the two instruments and the difference in the light travel time, by considering the differential rotation of the Sun \citep[method from][also implemented in SunPy]{Howard1990Rotation}. 

To improve data alignment and correct for any shift originating from small inaccuracies in our knowledge of the observing geometry (described in World Coordinate System, coordinates, see \citealt{Thompson2006WCS}), we derive and apply a preliminary distortion model to the re-projected data.
This distortion model is based on the pixel to pixel local correlation of the \blosOverMu{} by \sophi{} and the re-projected measurements of it by \hmi{} in the overlapping area. From the cross-correlation values, we derive a map giving the correct position of each pixel. Such an empirical method is susceptible to errors due to noise. Therefore, to minimise these errors, we fitted a second order polynomial surface to the resulting map. We apply the derived distortion model to the \hmi{} data after re-projecting it, instead of correcting the distortion in the \sophi{} measurements. This decision was taken here to preserve the intensity contrast observed at the limb, as methods readily available to correct the \sophi{} data, based on interpolations, lower it. 
To illustrate the quality of the data alignment, Fig.~\ref{Fig:Alignment} in Appendix \ref{App:B}
shows an example of a co-aligned \sophi{} and \hmi{} region together with cross-sections of \blosOverMu{} through the region.

\begin{figure*}[tbp]
   \sidecaption
    \begin{minipage}[c]{12cm}
    \centering
    \includegraphics[width=\hsize]{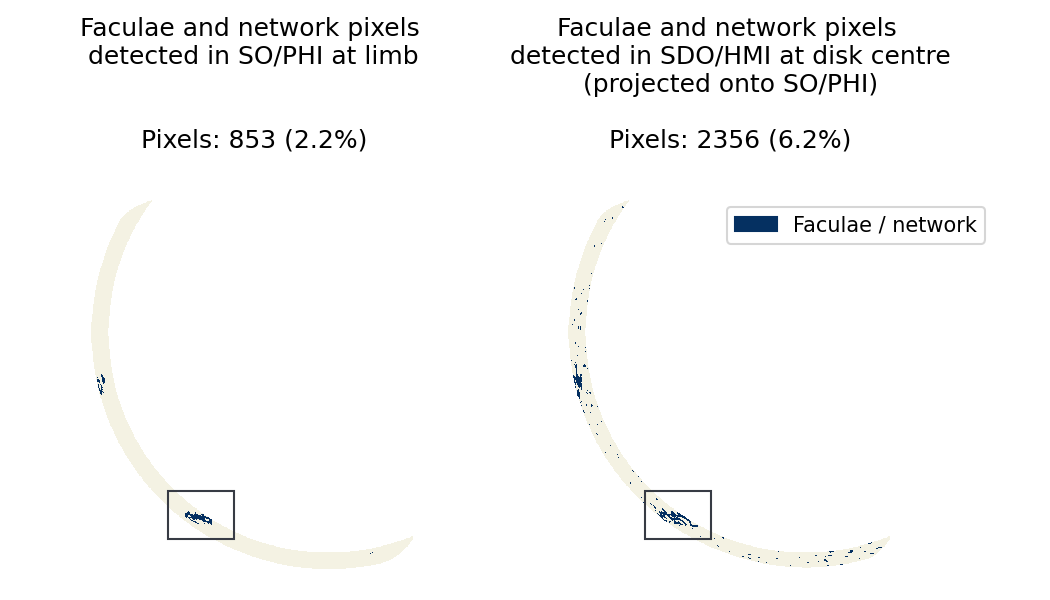}\\
    \includegraphics[width=\hsize]{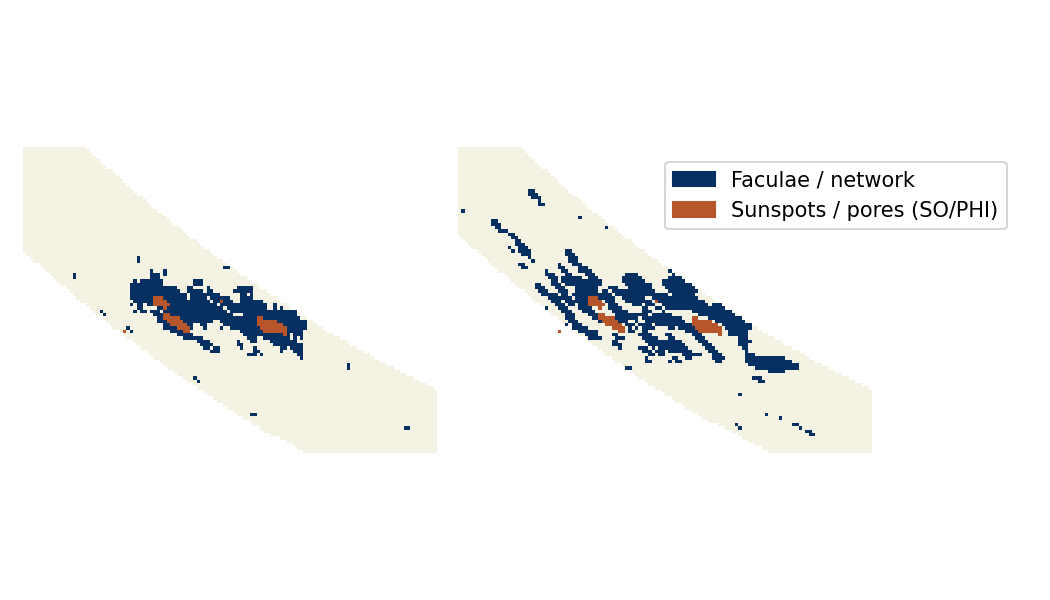}
    \end{minipage}
      \caption{Faculae map identified in \sophi{} (left) and in \hmi{}, re-projected to the coordinate system of \sophi{} (right). The upper row shows the area analysed in the study, the lower row shows a magnification of the area marked by the black rectangle on the top. In the magnified view, we also show the area that was identified as sunspot in the \sophi{} observations. We indicate the number of pixels identified as faculae in each figure (853 in \sophi{}, and 2356 in \hmi{}, representing $2.2$\,\% and $6.2$\,\% of the pixels in this area, respectively). The data shown here are from 6 Sept. 2021.
      }
         \label{Fig:FaculaeMap}
   \end{figure*}
   
We also re-project the facular map found in the \hmi{} magnetogram close to the disc centre (see Sect.~\ref{Sec:Identif}) with the method used for \blosOverMu{}, to \sophi{}'s coordinate system, and align the result with the distortion model described above.
This re-projected \hmi-based facular map consists of pixels that have varying amount of contribution from pixels identified as facular in the original \hmi{} data (before re-projection). 
In order to make sure that we analyse pixels that behave as faculae (i.e. the facular contribution is not insignificant compared to the internetwork contribution), we set a threshold based on \blosOverMu{}. We included in our analysis only those re-projected pixels that have the resulting \blosOverMu{} above the $3\sigma$ noise level of the \sophi{}-FDT (in line with the identification process of faculae that we used earlier, see Eq. \ref{Eq:Segm_Blos}). 
This is a conservative threshold, and we might miss some faculae that could be considered, however for this study we prioritise the avoidance of false positives over the inclusion of all facular pixels.
At the same time, we can now consider even stand-alone pixels as correct identification, as they were observed in several \hmi{} pixels close to the disc centre. Furthermore, we exclude any pixels that have contributions from sunspots or pores, as the measured intensity would be strongly affected by these features. The resulting facular map shows significantly more facular pixels near the limb than the \sophi{}-FDT magnetogram. Figure~\ref{Fig:FaculaeMap} compares the faculae maps obtained at the limb using \sophi{} data and at large $\mu$ from \hmi{} data, projected in the figure onto \sophi{}'s coordinates. The $12041$ facular pixels, identified close to the disc centre in the resampled \hmi{} magnetograms (from Fig. \ref{Fig:SegmentationExample}), convert into $2356$ pixels at the limb. The facular features at the limb, found through the re-projection of the \hmi{} facular map, represent $6.2\%$ of all the pixels in this region, compared to $2.2\%$ for those obtained directly from the \sophi{} limb data. This increase in the percentage of pixels identified close to limb with the data from disc centre indicates that many faculae with small flux density have been systematically missed at the limb in previous studies. However, the re-projection of the \hmi{} facular map shows fewer faculae in the close surroundings of sunspots. This has to do with the extended magnetic canopies of sunspots. In the near-limb \sophi{} observations, the magnetic canopy of the sunspot appears particularly extended (particularly in \blos{}), leading to sunspots being surrounded by pixels displaying a high magnetic flux density. These pixels are falsely identified as faculae in the \sophi{} facular map, whereas in the \hmi{} facular maps the same region is located at high $\mu$, and the canopies produce at most a very weak \blos{} signal close to the sunspot, which is not mistaken for faculae.

\begin{figure*}[tbp]
   \centering
    \begin{minipage}[c]{\hsize}
    \centering
    
    \includegraphics[width=\hsize]{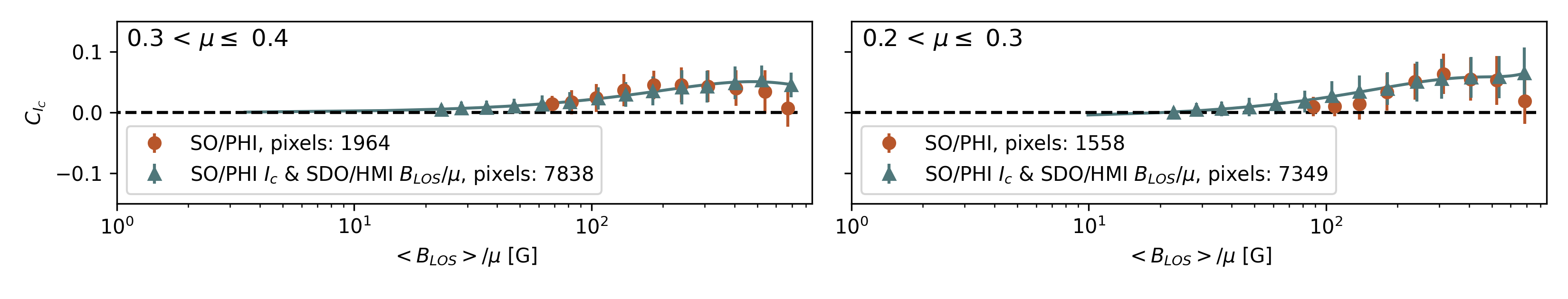}\\
    \hspace*{1mm}\includegraphics[width=.985\hsize]{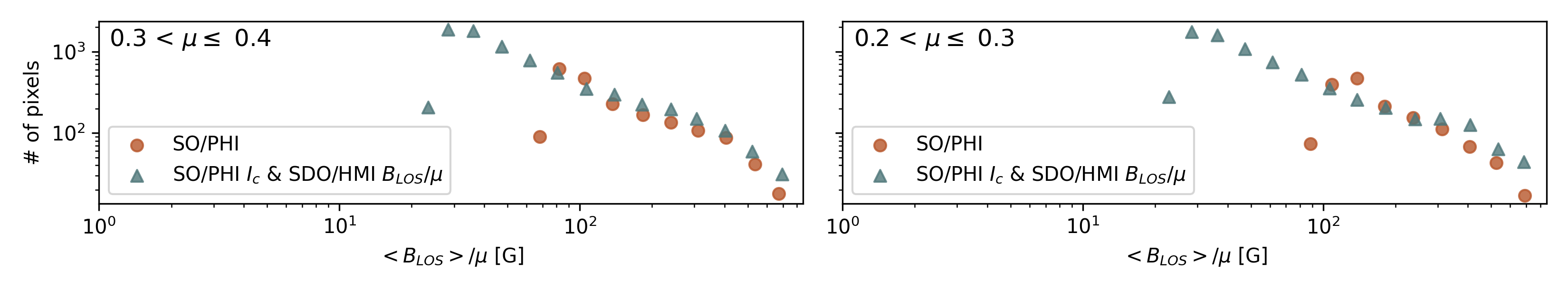}
    
    \end{minipage}

     \caption{Top: Facular contrast as a function of \blosOverMu{}, computed using the combined \sophi{} and \hmi{} data as well as the \sophi{} data alone in two $\mu$ intervals ($0.1 < \mu \leq 0.3$, right column, and $0.3 < \mu \leq 0.4$, left column). We split the data into equal intervals of $\log(B_{\rm LOS}/\mu)$. The error bars represent the standard deviation of the pixels within each bin. We fit a third order curve to the combined data points (continuous green line). Bottom: The number of pixels in each bin. We exclude from the plots \sophi{}-based data with $\mu < 0.2$.
     }
         \label{Fig:curvesNewMask_blos}
   \end{figure*}
   
\begin{figure*}
   \centering
   \begin{minipage}[c]{\hsize}
    \centering
    
    \includegraphics[width=\hsize]{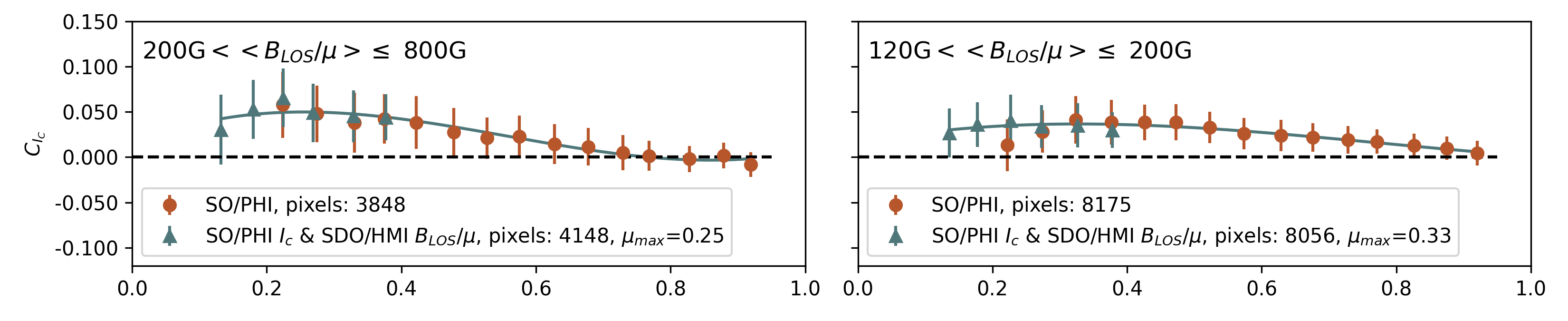}\\
    \includegraphics[width=\hsize]{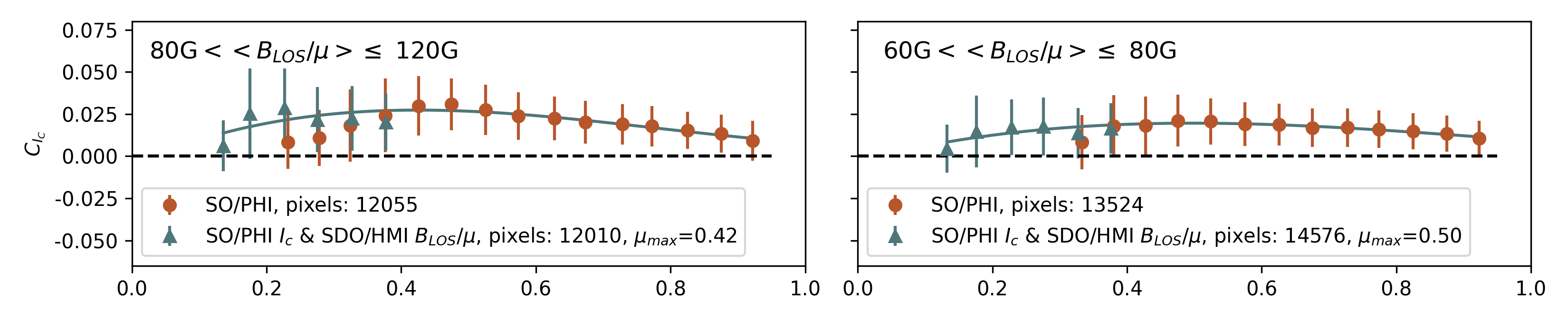}\\
    \includegraphics[width=\hsize]{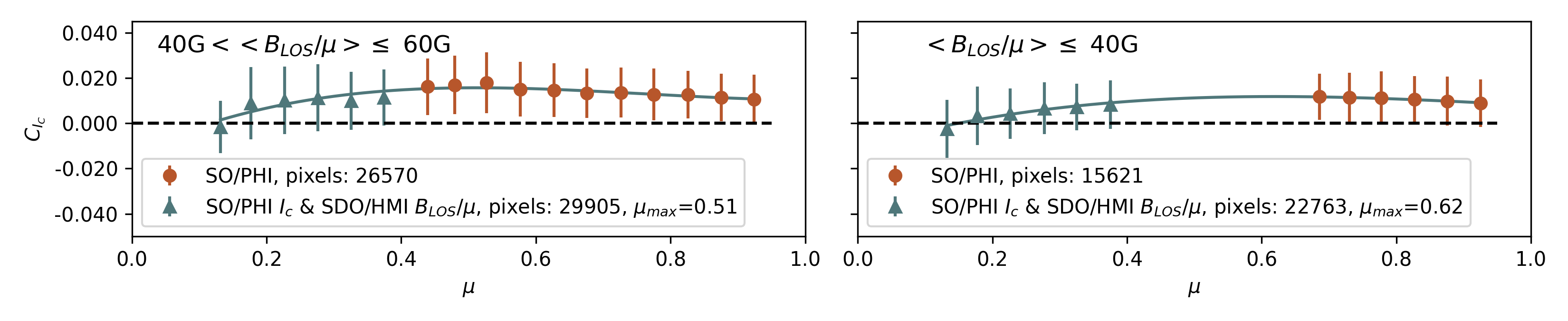}\\
    \hspace*{3mm}\includegraphics[width=.985\hsize]{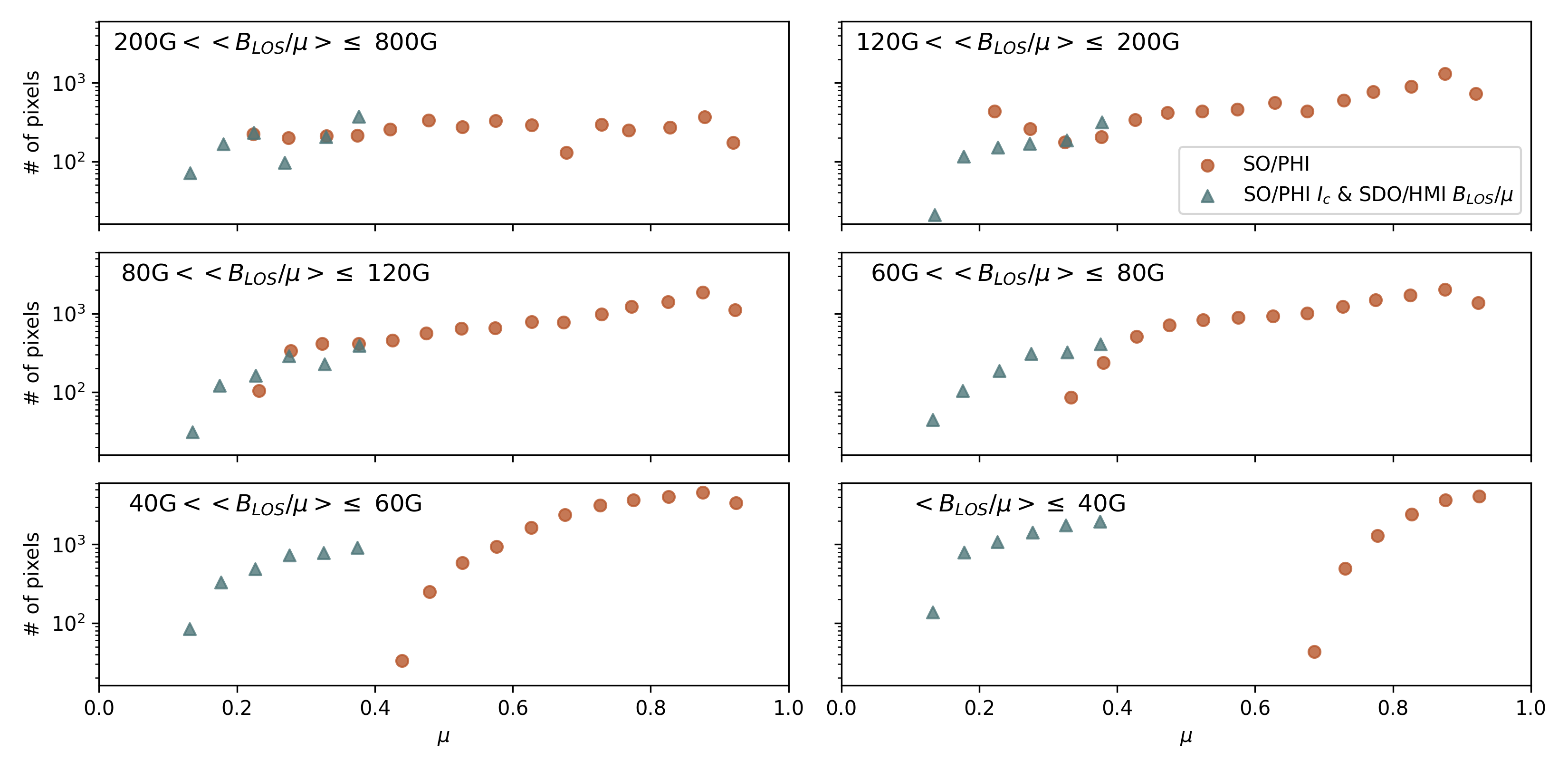}
    
    \end{minipage}

     \caption{Same as Fig.~\ref{Fig:curvesNewMask_blos}, but now showing facular contrast vs. $\mu$. The data are displayed in six intervals of \blosOverMu{}, and in bins of $\Delta\mu = 0.05$. The third order fit here is done through the combined points, extended with the points derived from \sophi{} only data where $\mu>0.4$.
     }
    \label{Fig:curvesNewMask_mu}
\end{figure*}

By inferring the \blos{} at low $\mu$-values from the re-projection of the same areas measured at a higher $\mu$, we reduce the uncertainty in the \blos{} values measured close to the limb.
We circumvent the problems arising at the limb, including the diminishing LOS component, lower light levels, increasing pixel coverage, changes in the formation height of the line, 
the radiative transfer effects, and the apparently extended sunspot canopy affecting faculae identification.

Our analysis involves a number of assumptions and approximations. Firstly, in our re-projection, we approximated the solar surface as a plane, ignoring any obstructions that occur in our line of sight due to the undulated solar surface. This could lead to cases where we identify faculae close to the disc centre, which are, however, obstructed by solar granulation closer to the limb. Such possible pixels would exhibit the intensity (and magnetic field) of internetwork at the limb, and, therefore, could bias the analysis towards lower contrast levels.
Secondly, we considered the studied flux tubes to be vertical with respect to the solar surface, and therefore assumed that a normalisation by $\mu$ accounts for the foreshortening.
Thirdly, we assumed that the different wavelength sampling of the spectral line, the somewhat different data reduction codes used to retrieve the \blos{}, and the different observation time (over $720$\,s for \hmi{}, and $33$\,s for \sophi{}) provide equivalent results. All these assumptions might have an effect on our results and should be considered further in future studies.

\section{Results and discussion}\label{Sec:Results}

The intensity contrast depends on both the location of the feature on the disc and the magnetic field strength averaged over the pixel. To disentangle the dependence of $C_{I_\text{c}}$ on each of these two factors, we consider individual $\mu$ and \blosOverMu{} intervals following \citet{ortiz2002Intensity}.

First, we consider the dependence on \blosOverMu{}. We first split all data into two $\mu$ intervals ($0.1<\mu\leq0.3$ and $0.3<\mu\leq0.4$) and look at the dependence of the contrast on \blosOverMu{} in each of them. In each interval, we bin the data into $14$ equal intervals of $\log(B_{\rm LOS}/\mu)$, and compute the bin-averaged $C_{I_\text{c}}$ (see Fig.~\ref{Fig:curvesNewMask_blos}). The standard deviation within the bins is shown in the figure as error bars. The two curves show the intensity contrast derived exclusively from \sophi{} data (brick-red markers), and from the combined measurements (i.e. $C_{I_\text{c}}$ from \sophi{} and \blos{} from \hmi{} re-projected to the coordinate system of \sophi{}, as described in Sect.~\ref{Sec:DataCombination}, shown in pine-green symbols). We show data with $\mu>0.1$ for the combined measurements, and with $\mu>0.2$ for \sophi{}-only data. The restriction to $\mu>0.2$ for these data is due to the low signal-to-noise ratio in the \sophi{} magnetograms at $\mu<=0.2$.
We plot the number of data points entering each $\left< B_{\rm LOS}\right>$ bin, separately, in the bottom panels.
When using the \sophi{} data alone, the weakest regions we could detect at $\mu <0.4$ were those with roughly \blosOverMu{} $\approx 80-90$\,G. By using the combined data, we could extend our analysis to regions as weak as \blosOverMu{} $\approx 20$\,G. 

Earlier studies found that, in regions not extremely near to the limb, the contrast of facular features initially increases with \blosOverMu{}, then usually decreases again at yet higher \blosOverMu{}-values \citep[see e.g.][]{ortiz2002Intensity,Yeo2013Intensity}.
Although the error-bars are rather high, we see a similar tendency in the top-left panel of Fig.~\ref{Fig:curvesNewMask_blos}, too.
At the same time, closer to the limb, the contrast of faculae keeps increasing with increasing magnetic flux density (see the top-right panel), also in agreement with previous studies.
This behaviour is due to the fact that for larger facular features (or pores) produced by stronger magnetic fields, we see increasingly more of the darker central part of the flux tube when approaching the disc centre, similarly to sunspots. Near the limb, we see more of the hot walls, so that all features are bright.

Similarly, we then split all the data into six \blosOverMu{} intervals and derive the dependence of the contrast on $\mu$ within each of these individual ranges (see Fig.~\ref{Fig:curvesNewMask_mu}). Our results suggest that studies involving two vantage points (such as this one) have the largest impact on pertaining to faculae with lower flux density. Therefore, we choose smaller intervals for the lower \blosOverMu{} range, and group a larger range into one interval for the higher flux densities. Another driver of this choice are the few magnetograms used in this study, which mean that there are relatively few pixels showing large flux densities.
Within each interval, we created bins that are $\Delta \mu = 0.05$ wide and calculate the mean intensity contrast in each (for reference, the widest pixel in the analysed area covers 0.02 $\mu$). Again, by combining data from two viewpoints (in pine-green), we significantly extend the $\mu$-range beyond what was possible with data from a single perspective (in brick-red), especially at low \blosOverMu{}.

The determination of the distance from disc centre, where the facular contrast is highest ($\mu_{\rm max}$), has been a long-standing debate in facular studies \citep[see][]{solanki_smallscale_1993}. To determine the $\mu_{\rm max}$ in our contrast curves, we fit a third order polynomial to the bins calculated from the combined \sophi{} and \hmi{} observations at $\mu \leq 0.4$, and from \sophi{} only data at $\mu > 0.4$ (see Fig.~\ref{Fig:curvesNewMask_mu}).
We calculated $\mu_{\rm max}$ based on the obtained fit, and compare our values to those reported in earlier studies by \citet{ortiz2002Intensity} and \citet{Yeo2013Intensity} in Table~\ref{table:mumax}. We find the trends in $\mu_{\rm max}$ with changing \blosOverMu{} to be similar to what earlier studies observed, which is the increase of $\mu_{\rm max}$ with decreasing magnetic flux density. However, we do not observe the inversion of this trend for the weakest flux densities. 
Our $\mu_{\rm max}$ values are also on average somewhat lower than those found by \citet{ortiz2002Intensity} and \citet{Yeo2013Intensity}.
We must consider, however, that the resolution of our data is lower than of that used in the compared studies. Another difference to these studies is the \blosOverMu{} intervals in which we derive $\mu_{\rm max}$. We note, however, that repeating the analysis for the intervals used by \cite{Yeo2013Intensity} and \cite{ortiz2002Intensity}, yielded no significant difference in the results.

\begin{table}[tb]
\caption{\label{table:mumax} Values of $\mu_{\rm max}$ calculated for various \blosOverMu{} ranges in this work, compared to what \cite{Yeo2013Intensity} and \citet{ortiz2002Intensity} found.}
\centering
\begin{tabular}{lp{50pt}p{50pt}p{50pt}}
\toprule[1.5pt]
\blosOverMu{} [G] & This study & \citet{Yeo2013Intensity} & \citet{ortiz2002Intensity}\\
\midrule[1.5pt]
$< 40$          & $0.62$ & & \\
$< 50$          & & $0.38$ & $0.50$\\
$40$ -- $60$    & $0.51$ & & \\
$50$ -- $80$    & & & $0.55$ \\
$60$ -- $80$    & $0.50$ & & \\
$50$ -- $100$   & & $0.45$ & \\
$80$ -- $120$   & $0.42$ & & $0.50$ \\
$100$ -- $180$  & & $0.45$ & \\
$120$ -- $200$  & $0.33$ & & $0.49$\\
$180$ -- $280$  & & $0.40$ & \\
$200$ -- $300$  & & & $0.42$ \\
$280$ -- $380$  & & $0.38$ & \\
$300$ -- $400$  & & & $0.42$ \\
$380$ -- $500$  & & $0.36$ & \\
$400$ -- $500$  & & & $0.35$ \\
$200$ -- $800$  & $0.25$ & & \\
$500$ -- $600$  & & & $0.22$ \\
$500$ -- $640$  & & $0.33$ & \\
$640$ -- $800$  & & $0.29$ & \\
\bottomrule[1.5pt]
\end{tabular}
\end{table}

Our extension of the observed $\mu$-range for low \blosOverMu{} values compared to the \sophi{}-only contrast curves indicates that combining two vantage points leads to more accurate $\mu_{\rm max}$ in such regions than enabled by single viewpoint observations. To this end, however, more and higher spatial resolution data are needed to be analysed than what is considered in the present study. Due to the large error bars in our results, which are the consequence of the provisional nature of the data, their low resolution,
the low statistics (only ten days of observations), and possible biases related to the normalisation of the \ic{} as a consequence of the low resolution and distortion, our results must be considered preliminary.

\section{Conclusions and outlook}\label{Sec:Conclusions}

One of the unprecedented opportunities offered by SO is that by co-observing together with other spacecraft in Earth-orbit, we have the possibility to observe the properties of a given region on the solar surface from two different vantage points. In this work, we use co-observations of \hmi{} and \sophi{}, with an approximately $60^\circ$ angle between their LOS, to study the dependence of facular brightness on the magnetic field strength and $\mu$ close to the solar limb. Earlier such studies faced the problem of strongly reduced magnetogram signals when measuring the LOS component of the magnetic field close to the limb. This is because of two main effects. Firstly, most of the field emerging in facular and network regions is roughly aligned with the solar surface normal, so that towards the limb its LOS component becomes increasingly weak, eventually falling below the instrumental noise level. Secondly, due to the reduced spatial resolution at the limb produced by foreshortening, the in-pixel averaging and cancellation effects of the measured flux density become more important. The combination of these effects, together with others (e.g. increased noise near the limb and the apparent extension of the magnetic canopy of sunspots), make it harder to detect faculae via their magnetic signature near the solar limb. Therefore, to measure the continuum intensity contrast in facular regions close to the limb, we combined \sophi{} continuum intensity measured close to the solar limb with the magnetic field co-observed by \hmi{} closer to the disc centre. Based on these data, we derived curves that describe the intensity contrast of facular pixels in relation to their position on the solar disc (expressed in $\mu$) and to their magnetic flux density (observed as \blos{}, and normalised to $\mu$ to counteract geometrical effects).

The preliminary results presented here highlight the potential of combining data from different angles. 
In particular, such a combined approach allows more reliable \blosOverMu{} measurements for areas with $0.1 < \mu \leq 0.4$.
As a consequence, we could identify and analyse faculae near the limb with significantly lower \blosOverMu{} values than possible from a single vantage point (e.g. that of \sophi{}), and thus extend the facular contrast curves to lower \blosOverMu{} and $\mu$ values. This allowed us to include in our analysis the $\mu$-ranges where the maximum of the contrast curves occurs ($\mu_{\rm max}$, i.e. the position of the turning point of the curves, where the contrast changes its trend from increasing towards the limb to decreasing) even for low \blosOverMu{} values.

Our results mostly confirm the trend of $\mu_{\rm max}$ observed by \cite{Yeo2013Intensity} and \cite{ortiz2002Intensity}, in that, apart from the lowest flux densities, $\mu_{\rm max}$ increases with decreasing \blosOverMu{}. For these ranges we find that $\mu_{\rm max}$ might lie closer to the limb than observations from a single point of view (e.g. that of \sophi{}, or of \hmi{}, see \citealt{Yeo2013Intensity}, \citealt{ortiz2002Intensity}) indicate. However, the same studies also observed an inversion of this trend around \blosOverMu$ \approx 50$\,G, which we cannot confirm.

Studies historically disagree on $\mu_{\rm max}$ \citep[see, e.g.][]{solanki_smallscale_1993} due to several factors, including (but not limited to), the resolution of the observations, the method of identification of facular features \citep[see][]{Centrone2003CLVFac_ident} and the systematic exclusion of features that could not be clearly identified as faculae \citep[see][]{AuffretMuller1991CLV_NBP}. 
The $\mu_{\rm max}$ values computed by us still have significant uncertainties, as our study is preliminary and can be improved in multiple aspects listed below.
However, the presented results suggest that analysing more and higher resolution data from two vantage points can lead to a more certain determination of $\mu_{\rm max}$ than possible from a single view point.

To consolidate the results presented here, different aspects of the present study need to be improved.
\begin{enumerate}
\item One obvious task is to  extend the investigation using combined data also beyond $\mu=0.4$, e.g. to close the gap in data points between $\mu=0.4$ and 0.8 seen in Fig.~\ref{Fig:curvesNewMask_mu} for $\left<B_{\rm LOS}/\mu\right> \leq 40$\,G. This is straightforward with increasing amount of available data, although it should be noted that the advantage of combining two vantage points decreases with increasing $\mu$.
\item Also, improving the statistics by including more data from \sophi{} would make the results more robust. Fortunately, SO is still in a relatively early phase of its science mission, holding the promise of many more observation campaigns from various angles between the spacecraft, the Sun, and Earth. Therefore, we expect to significantly improve the statistics compared with the present paper.
\item \sophi{} data with higher resolution would be of great value to overcome the problem of the large pixels and the poor resolution of magnetic features (not just close to the limb). This can be achieved either by employing data acquired at smaller distances from the Sun with the \sophi{}-FDT (which also provides an opportunity for a direct study on the effect of changing plate scale), or by using data from the \sophi{}-HRT, which will allow observing at even higher resolution than SDO/HMI, especially when close to perihelion. Such data (from both \sophi{} telescopes) will be particularly valuable after deconvolution of the PSF \citep[determined using phase diversity, see][]{Kahil2023_PSF, Bailen2023PSF}. For the impact of PSF reconstruction on facular studies with SDO/HMI, see \citet{Yeo2014PSF, Criscuoli2017PhotometryNetwFac_HMI_PSF}.
\item The \sophi{} data employed here have been reduced onboard preliminarily. The use of data reduced with improved methods is imperative. The data reduction methods of \sophi{} have already been refined beyond what was available at the time of the processing of the data for this study, and are being continuously improved further. 
\end{enumerate}
These improvements will be implemented in future studies.

%
%

\begin{acknowledgements}
We thank the referee for their insightful comments, that helped us to improve the paper. We are grateful to Kok Leng Yeo for her strong support and contribution to the work presented here. We thank Yang Liu for his support in investigating the SDO/HMI data products. This work has been carried out in the framework of the International Max Planck Research School (IMPRS) for Solar System Science at the Technical University of Braunschweig. Solar Orbiter is a space mission of international collaboration between ESA and NASA, operated by ESA. We are grateful to the ESA SOC and MOC teams for their support. The German contribution to SO/PHI is funded by the BMWi through DLR and by MPG central funds. The Spanish contribution is funded by AEI/MCIN/10.13039/501100011033/ and European Union “NextGenerationEU”/PRTR” (RTI2018-096886-C5,  PID2021-125325OB-C5,  PCI2022-135009-2, PCI2022-135029-2) and ERDF “A way of making Europe”; “Center of Excellence Severo Ochoa” awards to IAA-CSIC (SEV-2017-0709, CEX2021-001131-S); and a Ramón y Cajal fellowship awarded to DOS. The French contribution is funded by CNES. The SDO/HMI data are courtesy of NASA/SDO and the HMI Science Team.
\end{acknowledgements}

%
%
\bibliographystyle{aa}
\bibliography{bibfile.bib}

\appendix 
\onecolumn
\section{Re-projection of SDO/HMI data onto SO/PHI's coordinate system}\label{App:B}
   
\begin{figure*}[htbp]
   \centering
    \begin{minipage}[c]{\hsize}
    \centering
    
    \includegraphics[width=.59\hsize]{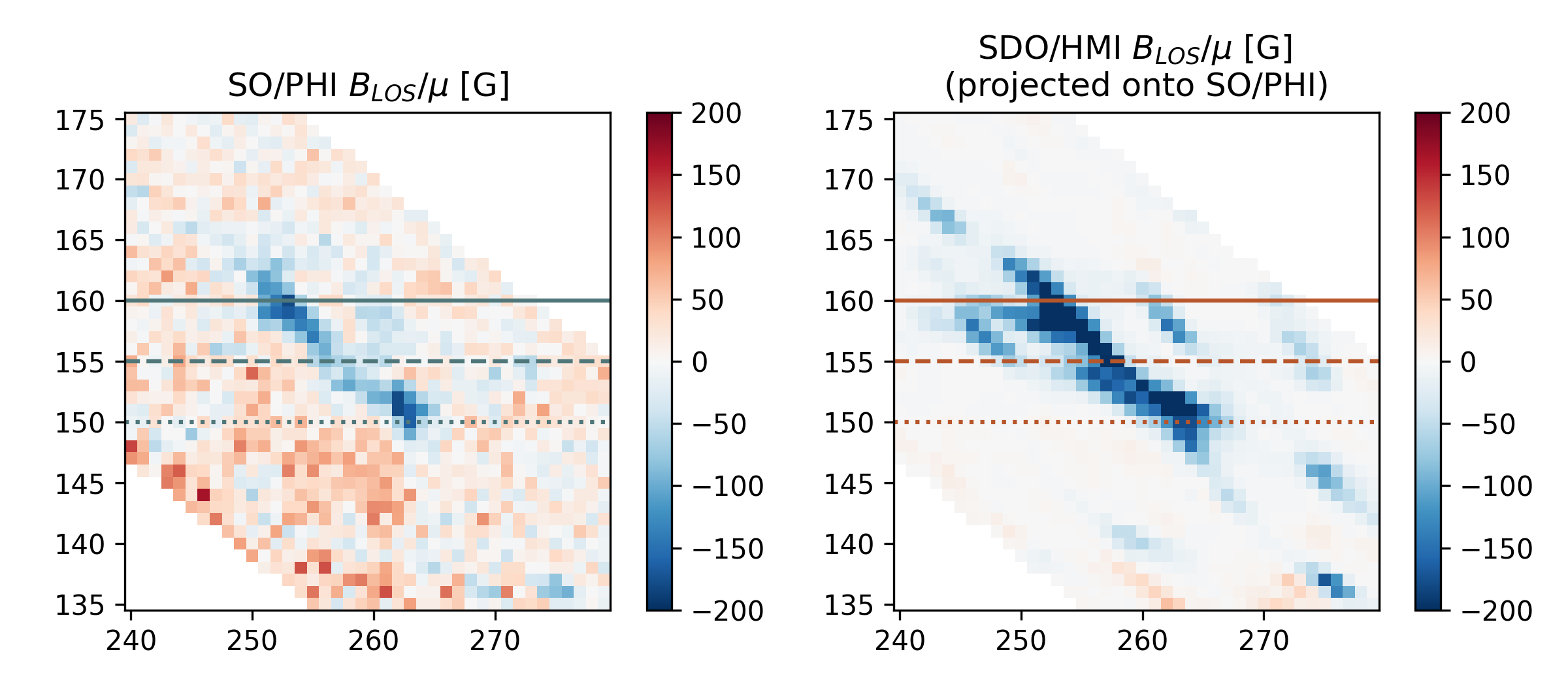}
    \includegraphics[width=.4\hsize]{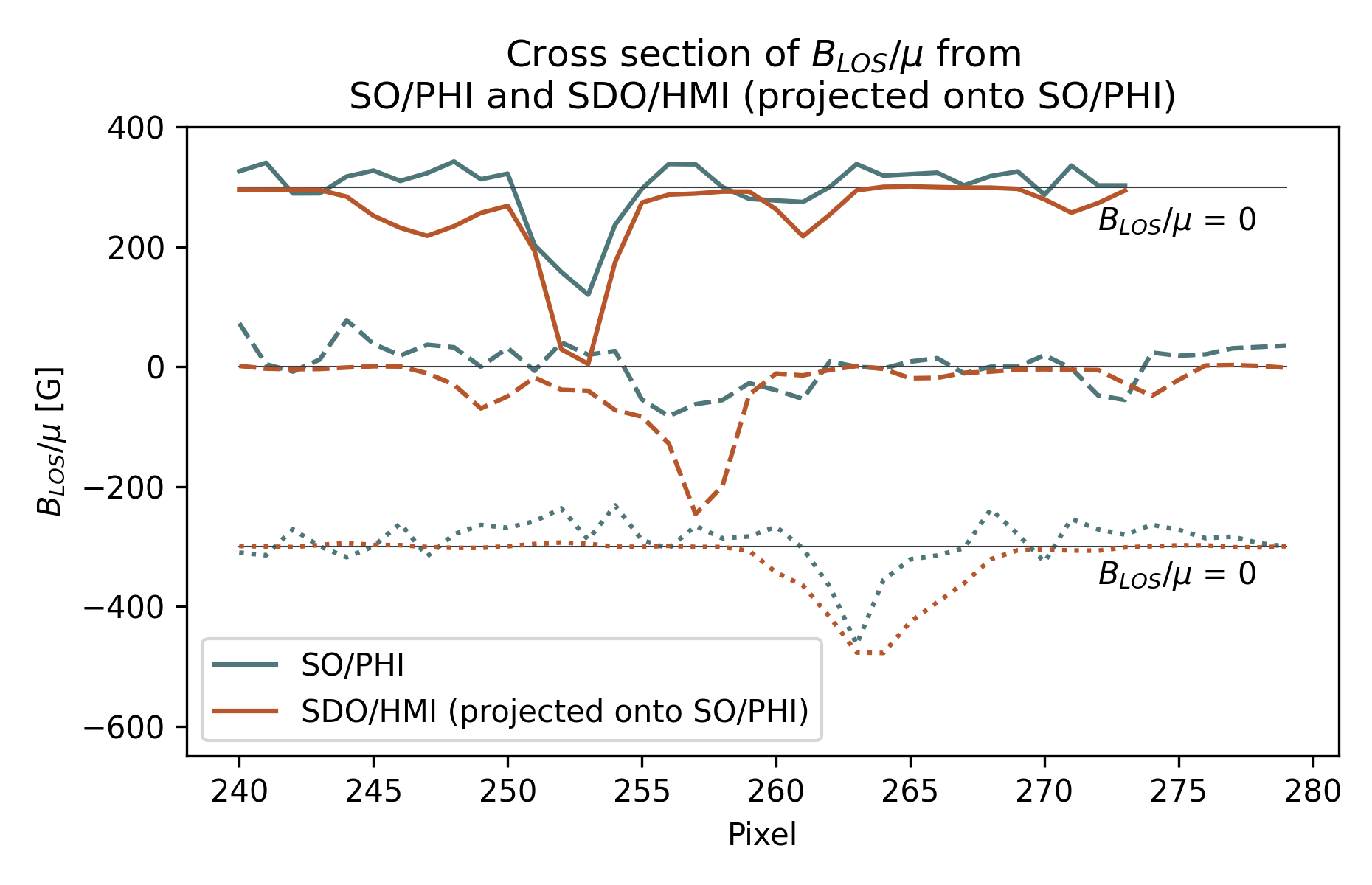}\\
    \includegraphics[width=.59\hsize]{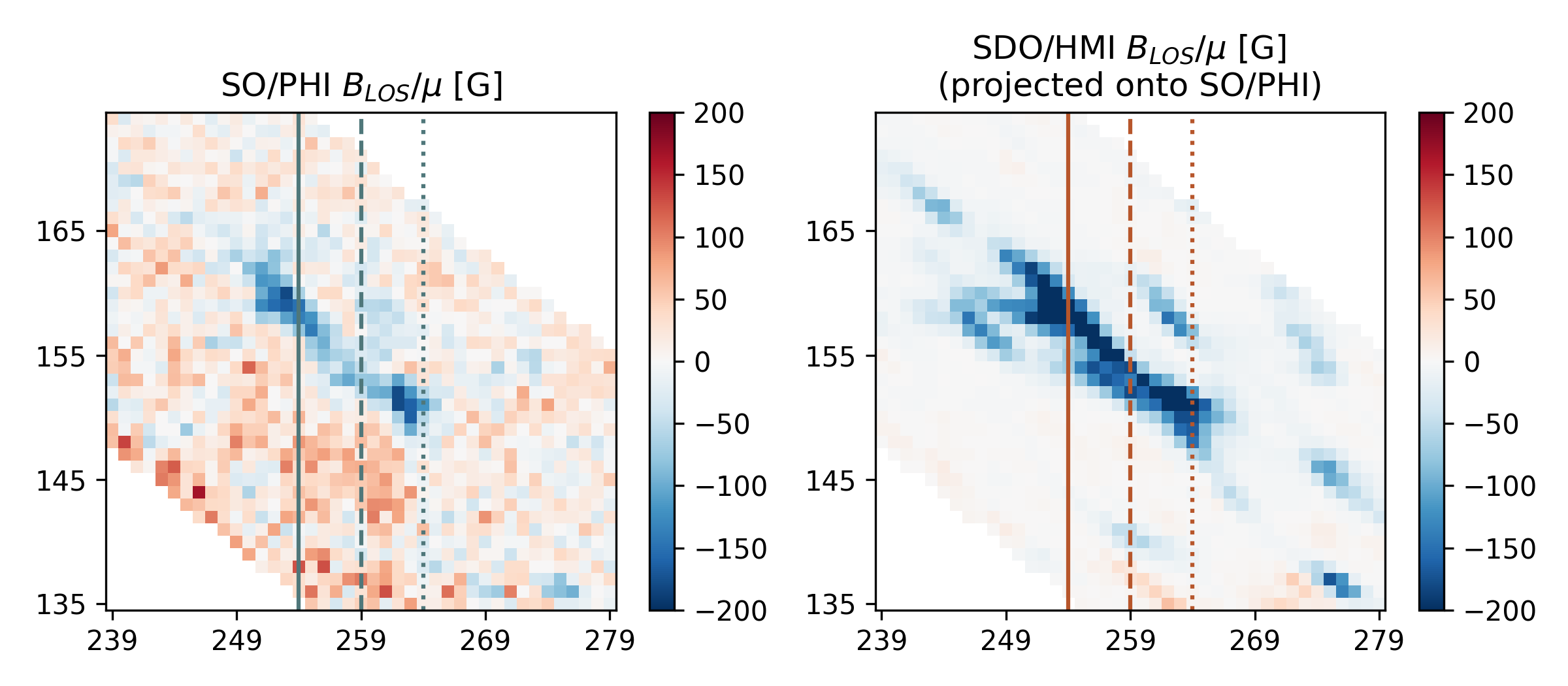}
    \includegraphics[width=.4\hsize]{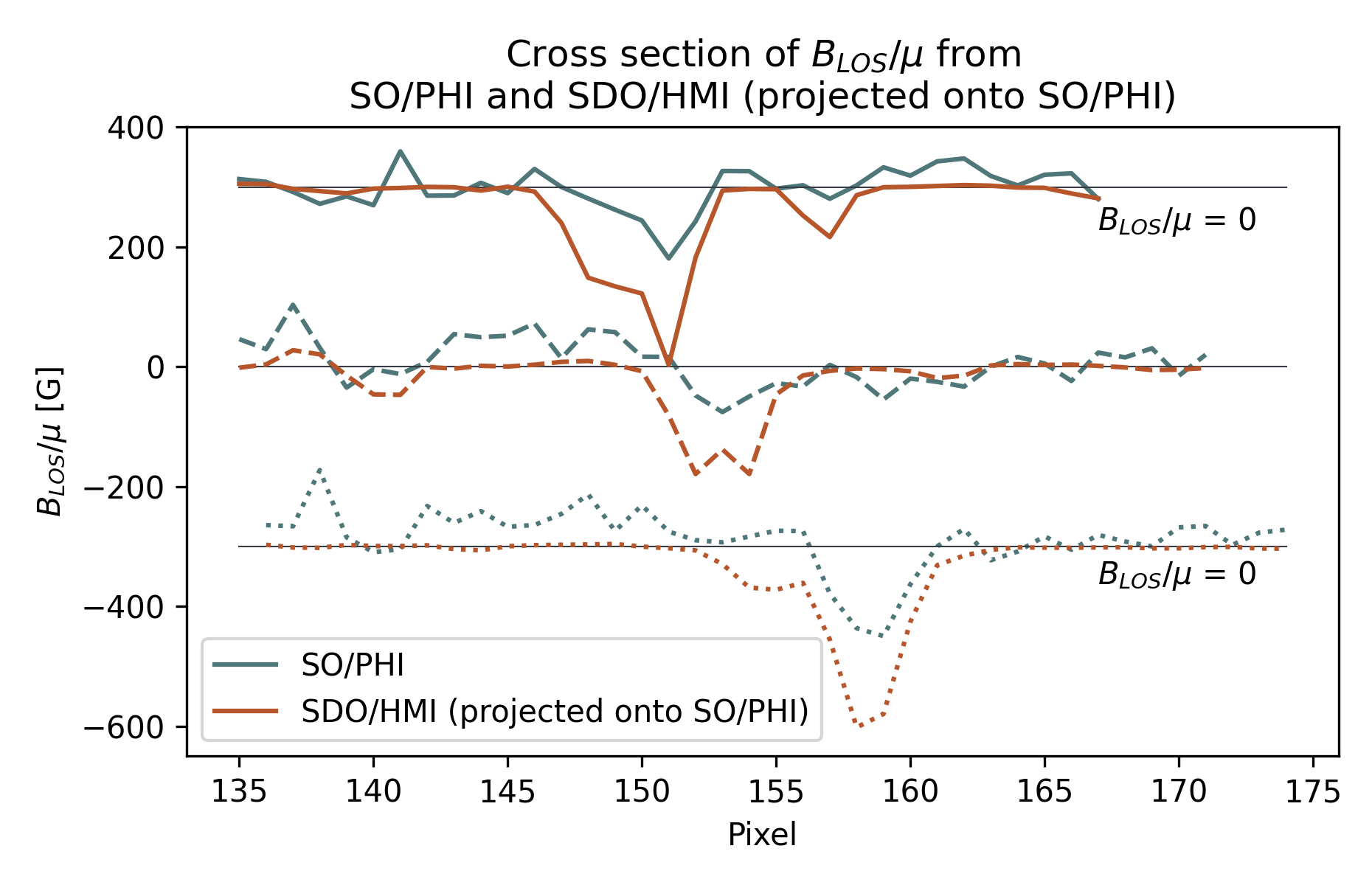}\\
    
    \end{minipage}

     \caption{A detailed view (left column) of \blosOverMu{} for a selected area in \sophi{} data at the limb and in \hmi{} close to disc centre, after re-projection to the limb. 
     The right column shows examples of \blosOverMu{} along cross-sections marked by the lines in the maps on the left. In the right panels, the different curves 
     are offset by $300$\,G in ordinate for better visibility (the respective $0$\,G is marked by the horizontal lines).}
         \label{Fig:Alignment}
   \end{figure*}

Figure \ref{Fig:Alignment} shows \blosOverMu{} over a selected part of the disc in both \sophi{} and \hmi{} data, to illustrate the quality of the data alignment of magnetic features selected from the solar scene. The \hmi{} data has been re-projected from the disc centre (where it was observed) to the coordinate system of \sophi{}, and aligned through local correlation of \blosOverMu{}, as described in Sect.~\ref{Sec:DataCombination}.
In the right panels, we show \blosOverMu{} along cross-sections marked by the corresponding lines on the maps (panels on the left). 
We note, that verifying the alignment is not straightforward due to the difference in the observation angle, as well as the smaller pixel size and lower noise levels in the re-projected \hmi{} observations.

Misalignment of the pixels would lead to assigning internetwork or sunspot/pore \ic{} pixels to facular \blosOverMu{} pixels. This would change the derived facular intensity contrast (by decreasing it and by shifting the curves in $\mu$ or \blosOverMu{}), as well as increase the standard deviation of the bins, shown in Figs.~\ref{Fig:curvesNewMask_blos} and \ref{Fig:curvesNewMask_mu}. Based on inspection of different areas, as shown in Fig.~\ref{Fig:Alignment}, we consider it unlikely that the misalignment of pixels is a major source of error in the contrast curves. However, a more robust distortion model allowing a more reliable alignment would certainly be of benefit for subsequent studies.

\FloatBarrier

\end{document}